\documentclass[aps,pre,notitlepage,reprint,longbibliography]{revtex4-1}

\usepackage{times}
\usepackage{amsmath}
\usepackage{amssymb}
\usepackage{bbold}
\usepackage{bm}
\usepackage{graphicx}
\usepackage{verbatim}
\usepackage{color}
\usepackage{subfigure}
\usepackage[pageanchor=true,
            plainpages=false,
            pdfpagelabels,
            bookmarks,
            bookmarksnumbered,
            colorlinks=true,
            allcolors=blue,
            pdfstartview=FitH]{hyperref}

\begin{document}

\title{Deterministic Interrelation Between Elastic Moduli in Critically Elastic Materials}

\author{Hongryol Jeon}
\affiliation{Department of Materials Science and Engineering, University of Illinois Urbana-Champaign, Urbana, IL 61801, USA}

\author{Mahdi Sadjadi}
\affiliation{Department of Physics,
Arizona State University, Tempe, AZ 85287, USA}

\author{Varda F. Hagh}
\email[To whom correspondence should be addressed:\\]
{hagh@illinois.edu}
\affiliation{Department of Mechanical Science and Engineering, and Beckman Institute for Advanced Science and Technology, University of Illinois Urbana-Champaign, Urbana, IL 61801, USA}

\begin{abstract}
Critically elastic materials--those that are rigid with a single state of self-stress--can be generated from parent systems with two states of self-stress by the removal of one of many constraints. We show that the elastic moduli of the resulting homogeneous and isotropic daughter systems are interrelated by a universal functional form parametrized by properties of the parent. In simulations of both spring networks and packings of soft spheres, judicious choice of parent systems and bond removal allows for the selection of a wide variety of moduli and Poisson's ratios in the critically elastic systems, providing a framework for versatile deterministic selection of mechanical properties. 


\end{abstract}
\maketitle


Disordered materials of homogeneous and isotropic structure are known to have exactly two independent linear elastic constants, expressible as a bulk modulus and a shear modulus~\cite{hodge1961isotropic, jog2006concise, gould1994introduction, slaughter2012linearized}. 
On the microscopic structural level, 
the rigidity and linear mechanical response of a disordered composite system, such as a packing of soft particles, can be described by the deformational response of an underlying elastic network, with each bond representing the interaction of two touching particles. A soft network with $N$ nodes and $N_c$ elastic bonds (often represented by springs) is minimally rigid (isostatic) if the number of constraining bonds $N_c$ balances the number of degrees of freedom ($Nd$ in $d$ dimensions)~\cite{maxwell1864calculation, calladine1978buckminster}. However, an isostatic system is not able to sustain generic stress due to its vanishing shear modulus~~\cite{ellenbroek2009non, wyart2008elasticity, zaccone2011approximate, zaccone2011network,ellenbroek2015rigidity}. For a network to be able to carry a force, it needs to have at least one state of self-stress, defined as independent modes of putting load on the local constraints (contacts in jamming and springs in the elastic network model) while keeping the entire system at static equilibrium. 
\begin{figure}[h!]
    \centering
    \includegraphics[width=8cm]{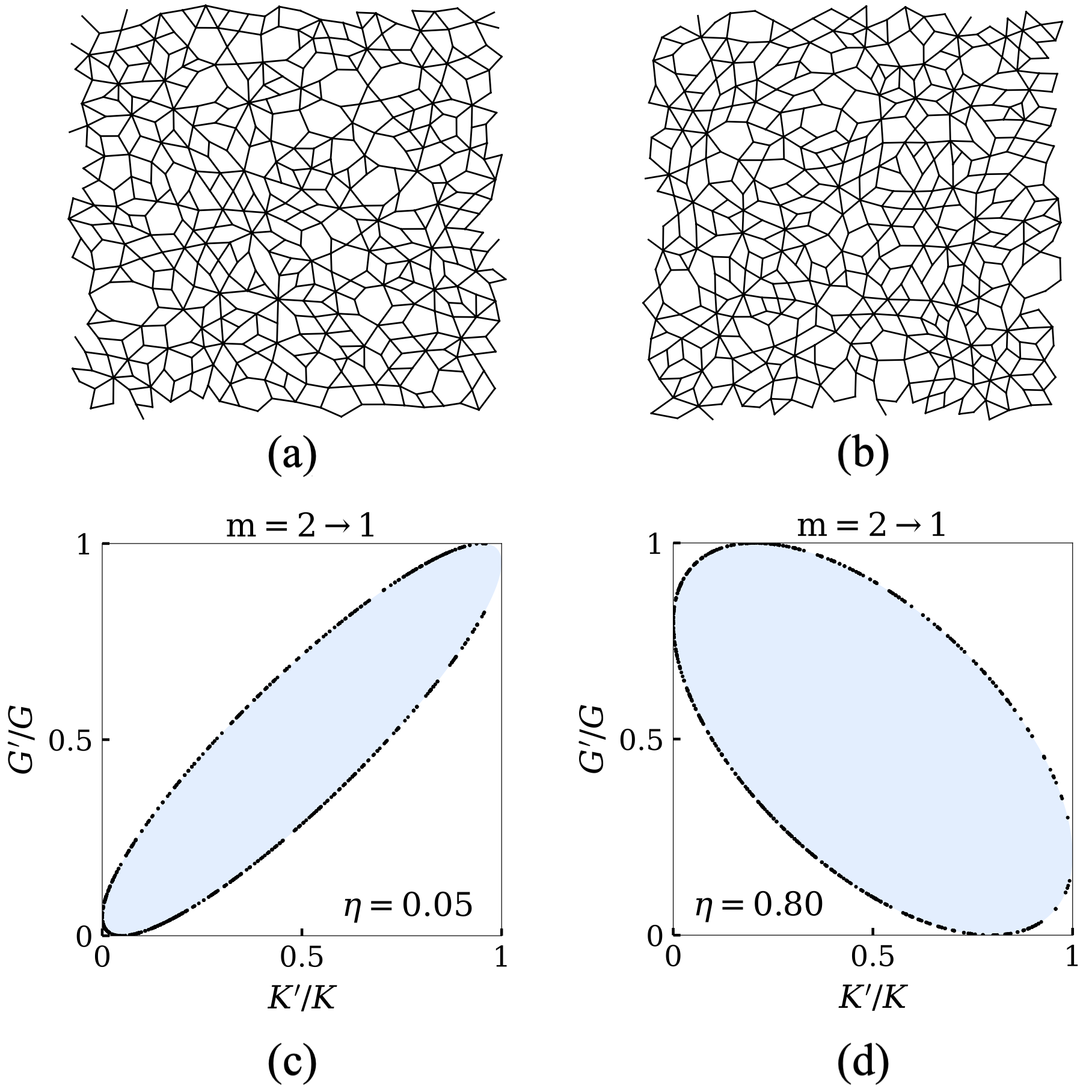}
    \caption{(a, b) Two parent spring networks with $m=2$ states of self-stress produced by randomly removing stressed bonds from a Delaunay triangulation while imposing Hilbert's stability condition on each node. (c, d) Values of the normalized bulk ($K'/K$) versus shear ($G'/G$) modulus after removing a single bond from networks in (a) and (b) respectively. The $m=2 \rightarrow 1$ nomenclature indicates that the elastic moduli are calculated for the daughter networks generated by removing a random constraint from the parent network. Each black data point corresponds to one daughter network thus generated.
    Both networks have similar structures but different correlations between the normalized moduli. The blue ellipses are drawn using Eq.~\ref{eq:ellipse}.   
    }
    \label{fig:network_ellipse}
\end{figure} 
Here, we focus on \textit{critically elastic} systems-- those that are rigid and support exactly one state of self-stress (SSS). We show that there is a deterministic functional dependence between the bulk and shear moduli of the critically elastic systems regardless of their preparation protocol, and that understanding this relation allows for the informed selection of a great variety of mechanical behaviors when these systems are generated from parents with two states of self-stress by removal of a single constraint. Fig.~\ref{fig:network_ellipse} illustrates this core finding in two spring network samples.

States of self-stress are described by a set of  orthogonal $N_c \times 1$  vectors, $\{ \mathbf{s}_i \}$, where the entries of each vector are the tensions in individual bonds. Without loss of generality, we will normalize these vectors to unit length. When a spring network with $m$ states of self-stress is deformed, the stored elastic energy is obtained from the projection of the deformation vector, $\bm{\Delta}$, onto the states of self-stress
\cite{lubensky2015phonons,hexner2018linking,paulose2015selective}:
\begin{align}
\delta E = \frac{1}{2} \sum_{i=1}^{m} (\mathbf{s}_i.\bm{\Delta})^2.
\label{eq:energy}
\end{align}
where $\bm{\Delta}$ is the $N_c \times 1$ vector of imposed changes in the individual bond lengths from their equilibrium lengths. 

All continuum mechanical behavior of a homogeneous and isotropic network can be obtained from two types of probing strains, often chosen as isotropic compressional strain, $\bm{\epsilon}_+$, and shear strain, $\bm{\epsilon}_{-}$. Here, we restrict ourselves to the $2D$ case where $(\bm{\epsilon}_{+})_{IJ} = \epsilon_{+} \delta_{IJ}$, and $(\bm{\epsilon}_{-})_{IJ} = (-1)^{I+1} \epsilon_{-} \delta_{IJ}$ where $\delta_{IJ}$ is the Kronecker delta function with $I,J=\{1,2\}$. However, the results presented are valid for any generic form of strain in $d$ dimensions, independent of the imposed boundary conditions. 
After obtaining the respective deformations $\bm{\Delta}_{+}$ and $\bm{\Delta}_{-}$ and their corresponding deformation energies, $\delta E_+, \delta E_-$, one can infer the elastic bulk and shear moduli, $K=\frac{\delta E_{+}}{2 V \epsilon_+^2}$ and $G=\frac{\delta E_{-}}{2 V \epsilon_-^2}$, where $V$ is the volume of the system. In the linear elastic limit, the modulus values are independent of $\epsilon_+, \epsilon_-$~\cite{gould1994introduction, slaughter2012linearized, hagh2018rigidity, hagh2022rigidpy}.
See Appendix~\ref{bulkshearstrain} for a more thorough discussion.



Removing constraints from an elastic network structure changes its moduli, which is key in many design algorithms for mechanical metamaterials. For instance, in tuning by pruning, one can tune a single elastic modulus (or the ratio of the two independent elastic moduli) in a predictable way by selective removal of constraints in multiple steps~\cite{goodrich2015principle, hagh2018disordered, reid2018auxetic, hagh2019broader, hexner2018linking}. When a system is far from marginal rigidity, meaning that there are more constraints than degrees of freedom, $N_c \gg Nd$, removing a single constraint typically does not change the elastic moduli significantly. Closer to the onset of rigidity, however, single constraint alterations can change the elastic moduli much more effectively, and as we show below, it reduces the number of independent elastic constants to one in going from $m=2\to 1$ SSS.

Note that in packings or networks with multiple states of self-stress, any superposition of these states is also a valid self-stress. In addition, removing any of the constraining bonds in areas of a \textit{parent} network where stress percolates (we call these stressed or load-bearing bonds as they will undergo stress if the network is deformed) removes one state of self-stress, transitioning the network from, say, $m$ SSS and $N_c$ bonds to a new \textit{daughter} network with $(m-1)$ SSS and $N_c-1$ bonds (denoted as $m \rightarrow m-1$). Instead of writing the daughter network's states of self-stress as vectors of dimension $N_c-1$, we can write them as $N_c$-dimensional vectors with a zero entry at the position of the removed bond in the parent network. This can be done at any given number of constraints above isostaticity.
The states of self-stress in the daughter system (denoted by \textit{primed} vectors) can thus be written as a linear combination of the states of self-stress in the parent system,
\begin{equation}
    \mathbf{s}'_j = \sum_{i=1}^{m} \alpha_{ij} \mathbf{s}_i\,
    \label{eq:newToOldSSS}
\end{equation}
where $\alpha_{ij}$ represent $m(m-1)$ coefficients.
However, not all $\alpha_{ij}$ are independent. Consistent with the treatment of $\mathbf{s}_i$ in the parent network, the states of self-stress in the daughter network are again orthonormal ($\mathbf{s}'_i \cdot \mathbf{s}'_j = \delta_{ij}$). This adds $m(m-1)/2$ constraints to the set of new states of self-stress. Furthermore, the tension on the removed bond (say, bond $l$) is required to be zero ($s'_{j,l} = 0$), which leaves the set of self-stresses with another $m-1$ constraints. Thus, the total number of independent free parameters, $\alpha_{ij}$, is $(m-1)(m-2)/2$. While for large $m$ there are $\mathcal{O}(m^2)$ free parameters, neither of the $m=2$ or $m=1$ cases have any free parameters left. The case of $m=1$ is rather trivial since the system has no remaining states of self-stress once a constraint is removed. The $m=2$ case is non-trivial and the focus of the current work. According to the above analysis, the two states of self-stress in the $m=2$ parent network uniquely determine the single state of self-stress in the daughter network after a single bond is removed. 

Let $(\mathbf{s}_1, \mathbf{s}_2)$ be the states of self-stress of a $m=2$ parent network. Using Eq.~\ref{eq:newToOldSSS}, we can write the single self-stress of the daughter network, $\mathbf{s}'$, as:
\begin{equation}
    \mathbf{s}' = \frac{\mathbf{s}_1 + \chi_l \mathbf{s}_2}{\sqrt{1+\chi_l^2}}.
    \label{eq:sprime}
\end{equation}
Note that $\mathbf{s'}$ is normalized and $\chi_l$ must be chosen such that the tension on spring $l$ is zero, i.e. $s'_l = 0$. Thus $\chi_l = - {s_{1,l}}/{s_{2,l}}$ assuming $s_{2,l}\neq0$ (without loss of generality, as for the bond $l$ at least one of $s_{1,l},s_{2,l}$ is non-zero). We drop the subscript $l$ in the following steps as $l$ can be any of the bonds in stressed regions in the parent system. 
We will now use Eq.~\ref{eq:sprime} to write the moduli ($K', G'$) of the $m=1$ daughter network as functions of the $m=2$ parent network moduli, $K, G$.


Eq.~\ref{eq:energy} 
with $\bm{\Delta}_{+}$ and $\bm{\Delta}_{-}$ corresponding to the deformations of the two elastic moduli yields for the parent network:
\begin{align}
K &= \frac{1}{4} [ \left(\mathbf{s}_1.\bm{\Delta}_{+}\right)^2 + \left(\mathbf{s}_2.\bm{\Delta}_{+}\right)^2 ] = \frac{1}{4} (a_1^2 + a_2^2), \nonumber \\
G &= \frac{1}{4} [ \left(\mathbf{s}_1.\bm{\Delta}_{-}\right)^2 + \left(\mathbf{s}_2.\bm{\Delta}_{-}\right)^2 ] = \frac{1}{4} (b_1^2 + b_2^2),
\end{align}
where the second equalities in each equation define $a_1$, $a_2$, $b_1$, $b_2$. The analogous relations for the daughter network, together with  Eq.~\ref{eq:sprime}, give the elastic moduli after removing a bond:
\begin{align}
K' &= \frac{1}{4} (\mathbf{s}'.\bm{\Delta}^{'}_{+})^2 = \frac{[\left(\mathbf{s}_1 + \chi \mathbf{s}_2\right).\bm{\Delta}_{+}]^2}{4(1+\chi^2)} = \frac{\left(a_1 + \chi a_2\right)^2}{4(1+\chi^2)},  \nonumber \\
G' &= \frac{1}{4} (\mathbf{s}'.\bm{\Delta}^{'}_{-})^2 = \frac{[\left(\mathbf{s}_1 + \chi \mathbf{s}_2\right).\bm{\Delta}_{-}]^2}{4(1+\chi^2)} =  \frac{\left(b_1 + \chi b_2\right)^2}{4(1+\chi^2)}.
\end{align}

\begin{figure}[h!]
    \centering
    \includegraphics[width=8.4cm]{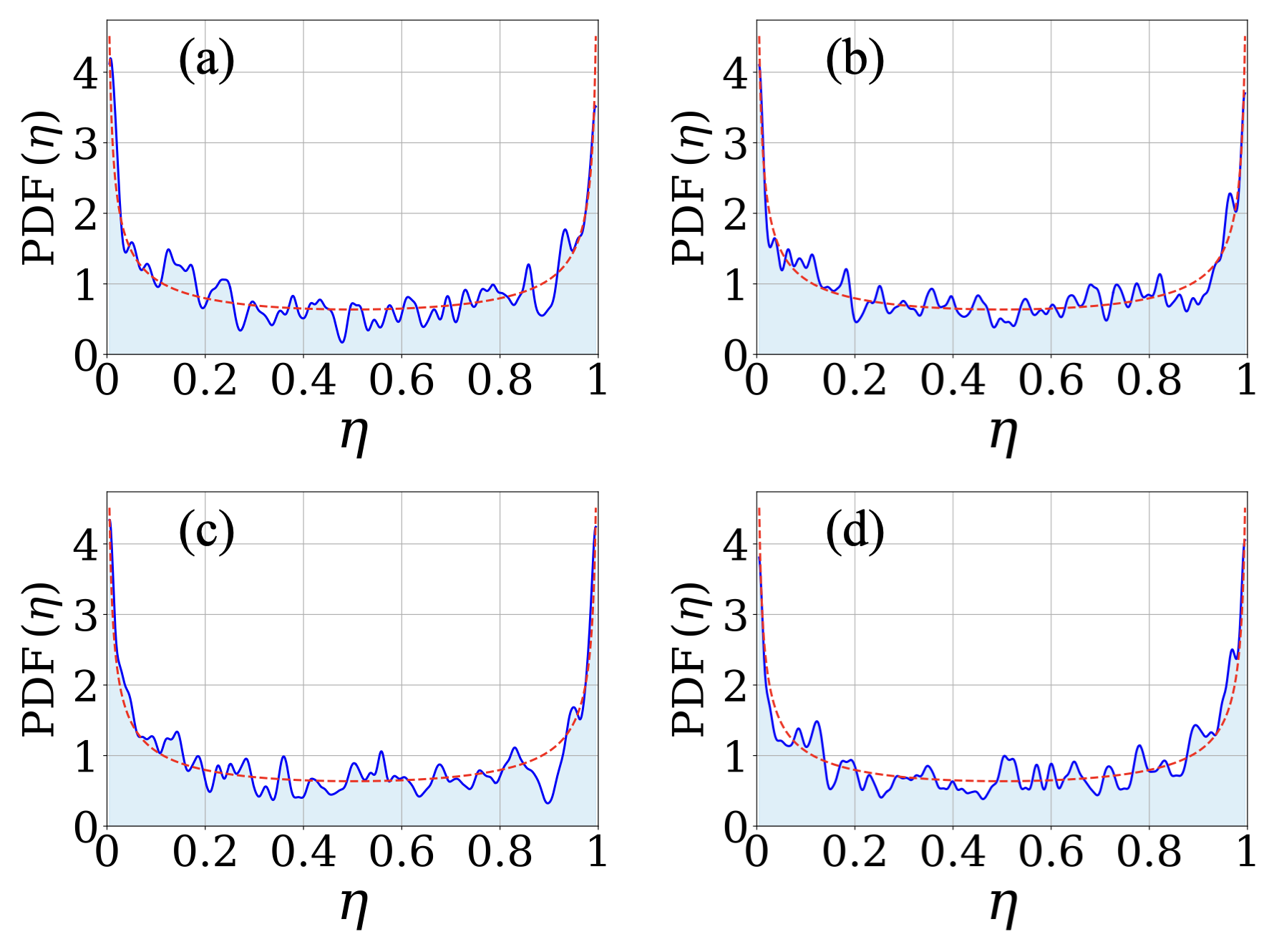}
    
    \caption{
    Probability density functions (PDFs) of $\eta$ estimated via Kernel Density Estimation (KDE), compared with the theoretical arcsine distribution 
    \( f(\eta) = \frac{1}{\pi \sqrt{\eta(1 - \eta)}}, \quad \eta \in (0, 1) \), under different random perturbations of magnitude $dr$ in two networks with $m=2$ SSS and different connectivity patterns, shown in Fig.~\ref{fig:network_ellipse}a,b. 
    Each panel shows the KDE-based estimate from $1000$ samples of $\eta$ (blue curve, bandwidth = $0.02$), overlaid with the analytical arcsine PDF (red dashed line).\\
    \textbf{(a)} $dr = 0.01$, from network in Fig.~\ref{fig:network_ellipse}a; \\
    \textbf{(b)} $dr = 0.005$, from network in Fig.~\ref{fig:network_ellipse}a; \\
    \textbf{(c)} $dr = 0.01$, from network in Fig.~\ref{fig:network_ellipse}b; \\
    \textbf{(d)} $dr = 0.005$, from network in Fig.~\ref{fig:network_ellipse}b.
    }
    \label{fig:distribution}
\end{figure}

It is evident that $(K', G')$ are bounded by the values of elastic moduli in the parent network, i.e. $0 \le K' \le K$ and $0 \le G' \le G$. Therefore, it is useful to define the normalized elastic moduli $K'/K,G'/G$ as:
\begin{align}
    \frac{K'}{K} &= \frac{(a_1+\chi a_2)^2}{(a_1^2+a_2^2)(1+\chi^2)}, \nonumber \\
    \frac{G'}{G} &= \frac{(b_1+\chi b_2)^2}{(b_1^2+b_2^2)(1+\chi^2)}.
\end{align}
These equations are the parameterization of an ellipse in terms of $\chi$ (see Appendix \ref{ellipse_eq_proof} for proof). This means the elastic moduli of the network with $m=1$ state of self-stress are not independent and are related through an elliptic equation which, in standard form, can be written as:
\begin{equation}\label{eq:ellipse}
    \left(\frac{\frac{K'}{K}+\frac{G'}{G}-1}{\sqrt{1-\eta}}\right)^2 + \left(\frac{\frac{K'}{K}-\frac{G'}{G}}{\sqrt{\eta}}\right)^2 = 1,
\end{equation}
where $\eta$ is:
\begin{equation}
    \eta \equiv \frac{(a_1 b_2 -  a_2 b_1)^2}{(b_1^2+b_2^2)(a_1^2+a_2^2)}.\
\label{eq:eta}
\end{equation}

In large systems with many available constraints, the distribution of $\chi$ covers a wide range of values and this parameterization puts all values of the moduli after removing a bond on the perimeter of an ellipse given by Eq.~\ref{eq:ellipse}. Two examples are shown in Fig.~\ref{fig:network_ellipse}, where the black data points on each ellipse correspond to the normalized elastic moduli $G'/G$ and $K'/K$ after removing a random bond from the corresponding parent network. Both networks in Fig.~\ref{fig:network_ellipse}a,b are prepared using a selective pruning algorithm in which stressed bonds are removed randomly from a Delaunay triangulation~\cite{lee1980two} while imposing Hilbert's stability condition,
requiring each node to have at least $d+1$ bonds, positioned such that any two adjacent bonds make an angle smaller than $\pi$~\cite{alexander1998amorphous, lopez2013jamming}.

Note that $\eta$ is determined by the elastic response of the parent network at $m=2$, and its value ranges from $0$ to $1$. Moreover, $\eta$ fully controls the shape of the ellipse.  Eq.~\ref{eq:ellipse} is the equation of an ellipse with its center at (1/2, 1/2). When $\eta \in (0, 1/2)$, the major axis of the ellipse is $\sqrt{1-\eta}$ and oriented along ${K'}/{K}-{G'}/{G}=0$ (under a $\left[+\frac{\pi}{4}\right]$ angle), while for $\eta \in (1/2, 1)$, the major axis is $\sqrt{\eta}$ and oriented along ${K'}/{K}+{G'}/{G}=1$ (under a $\left[-\frac{\pi}{4}\right]$ angle). The ellipse becomes a circle for $\eta=1/2$ and asymptotes to a line for $\eta=0$ or $\eta=1$.


The angle of the ellipse divides critically elastic systems into two classes (as demonstrated in Fig.~\ref{fig:network_ellipse}). In the first class, identified with a $\left[+\frac{\pi}{4}\right]$ angle, elastic moduli change in a positively correlated fashion, i.e., the system is robust against removal of certain bonds (where both bulk and shear moduli change slightly), but very sensitive against the removal of others (where both bulk and shear moduli are reduced dramatically).
The second class consists of those systems that produce a $\left[-\frac{\pi}{4}\right]$ angle ellipse. Here, changes in the bulk and shear modulus are negatively correlated: if removing a bond changes the shear modulus drastically, it leaves the bulk modulus virtually unchanged and vice versa. These networks are suitable when one intends to change only one of the elastic moduli significantly. 

Our studies on networks created with different protocols (e.g., jammed packings of soft harmonic particles or selective pruning of bonds from a Delaunay triangulation) show that regardless of the used protocol, resulting networks can have a positively or negatively correlated elastic response at $m = 1$. 
The quantity $\eta$ lacks a direct physical interpretation; however, since its value determines the class to which a network belongs, it is crucial to understand its dependence on underlying structural parameters. As shown in Fig.~\ref{fig:network_ellipse}, $\eta$ varies with changes in network connectivity and constraint patterns. Moreover, even for a fixed connectivity, $\eta$ is sensitive to the spatial configuration of the nodes. To investigate how positional perturbations influence $\eta$, we begin with the parent networks illustrated in Fig.~\ref{fig:network_ellipse} and apply random displacements of magnitude $dr = 0.005$ and $dr = 0.01$ to the node positions. The bonds are modeled as Hookean springs with uniform stiffness $k = 1$, and following each perturbation, we minimize the total potential energy, $U = \frac{1}{2} k \sum_{i,j} (\delta l_{i,j})^2$, where $\delta l_{i,j}$ is the change in length of the bond connecting nodes $i$ and $j$ relative to its equilibrium value. 

Fig.~\ref{fig:distribution} shows the resulting distribution of $\eta$ values obtained from $1000$ independent perturbations. The probability density function (PDF) of $\eta$ consistently exhibits a U-shaped profile, closely following the arcsine function, regardless of the initial connectivity or perturbation magnitude. Letting $\theta$ denote the angle between $\bm{\Delta}{+}$ and $\bm{\Delta}{-}$, Eq.~\ref{eq:eta} implies $\eta = \sin^2\theta$. Under random spatial perturbations, $\theta$ becomes approximately uniformly distributed over the interval $[0, \pi]$, resulting in an arcsine distribution for $\eta$ (see Appendix~\ref{Arcsine} for derivation), with PDF:
\begin{equation}
f(\eta) = \frac{1}{\pi \sqrt{\eta (1 - \eta)}}, \quad \text{for } \eta \in (0, 1).
\label{eq:pdf}
\end{equation}
This result demonstrates that $\eta$ can be tuned not only through network connectivity but also via the spatial distribution of nodes. It further implies that random sampling of the energy landscape in disordered networks with $m = 1$ state of self-stress and fixed connectivity is more likely to yield values of $\eta$ near $0$ or $1$. In other words, an elliptical relation between the two elastic moduli is statistically favored over a circular one.
\begin{figure*}[ht]
    \centering
    \includegraphics[scale=0.35]{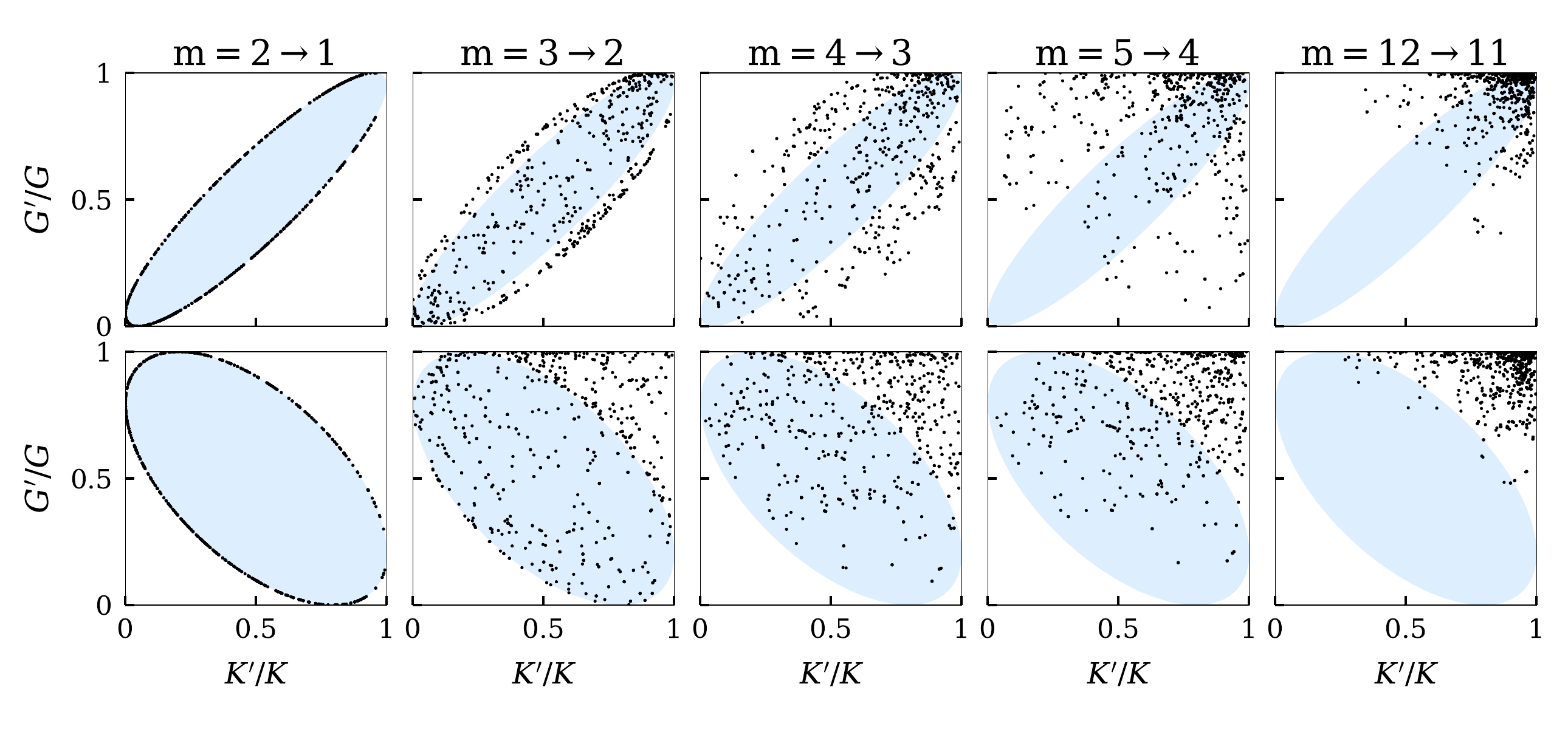}
    \caption{Normalized elastic moduli, $(K'/K, G'/G)$ for $m=2\to 1$, $3\to 2$, $4\to 3$, $5\to 4$ and $m=12 \to 11$ SSS in two sets of networks with $\left[\frac{\pi}{4}\right]$ (top) and $\left[-\frac{\pi}{4}\right]$ (bottom) orientations. For larger $m$, the networks exhibit more robust response to bond removal and changes in elastic moduli after removing a bond tend to be smaller, leading to the clustering of moduli values near upper right corner. However, closer to $m=1$, the elastic response can vary significantly by removing a single bond due to the criticality of the network. Although for $m=3\to 2$ the normalized elastic moduli are no longer on the perimeter of the characteristic ellipse, the signature of the ellipse is still present in the elastic response.}
    \label{fig:ellipse_evolution}
\end{figure*}

The change in elastic moduli upon transitioning from $m=2$ to the critically elastic, $m=1$ state, shows qualitative differences to transitions at higher numbers of states of self-stress. Fig.~\ref{fig:ellipse_evolution} compares the normalized elastic moduli of the $m=2\to 1$ transition (which produces the ellipse) to transitions at larger $m\to m-1$ for two sets of networks. Starting with $m=12$ SSS, we sequentially remove constraints to reach lower numbers of SSS. We then use the resulting networks as the parent network for each panel presented in the figure.
Although the details depend on the specifics of the network, we generally observe that the deterministic functional dependence between the bulk and shear moduli quickly disappears as we go to larger $m$. While the normalized moduli maintain some degree of elliptical signature when we go from $m=3\to 2$, this correlation is quickly lost for higher $m\to m-1$. It can also be seen from Fig.~\ref{fig:ellipse_evolution} that already for moderately large $m$, both $K'/K$ and $G'/G$ approach $1$ for most bond removals, which is intuitive as only exceptional bonds can significantly alter the mechanics of a network with many states of self-stress. Note, however, that for intermediate values such as $m=3,4$ the scattered outputs cover a greater area of the normalized moduli unit square, making these cases attractive if a wide variation of mechanical response is desired. 

Since normalized elastic moduli form  a closed curve, any horizontal or vertical line would cross the curve at exactly two points. This means when $N_c \rightarrow \infty$, for any given value of the bulk (shear) modulus, there are two bonds whose removal leads to different values for the shear (bulk) modulus, implying interrelated pairs of bonds.
This effect can be best represented by the Poisson's ratio, defined as:
\begin{equation}
    \nu = \frac{dK - 2G}{d(d-1)K + 2G}
    \label{eq:PoissonsRatio}
\end{equation}
for a generic $d$ dimensional elastic system. Note that Poisson's ratio is bounded by $-1<\nu<\frac{1}{d-1}$. For ordinary isotropic materials, Poisson's ratio is always positive. For systems with $m=1$ SSS in $2D$, Eq.~\ref{eq:PoissonsRatio} reduces to:
\begin{equation}
    \nu' = \frac{K'-G'}{K'+G'}.
\end{equation}

\begin{figure}[h!]
    \centering
    \includegraphics[width=\linewidth]{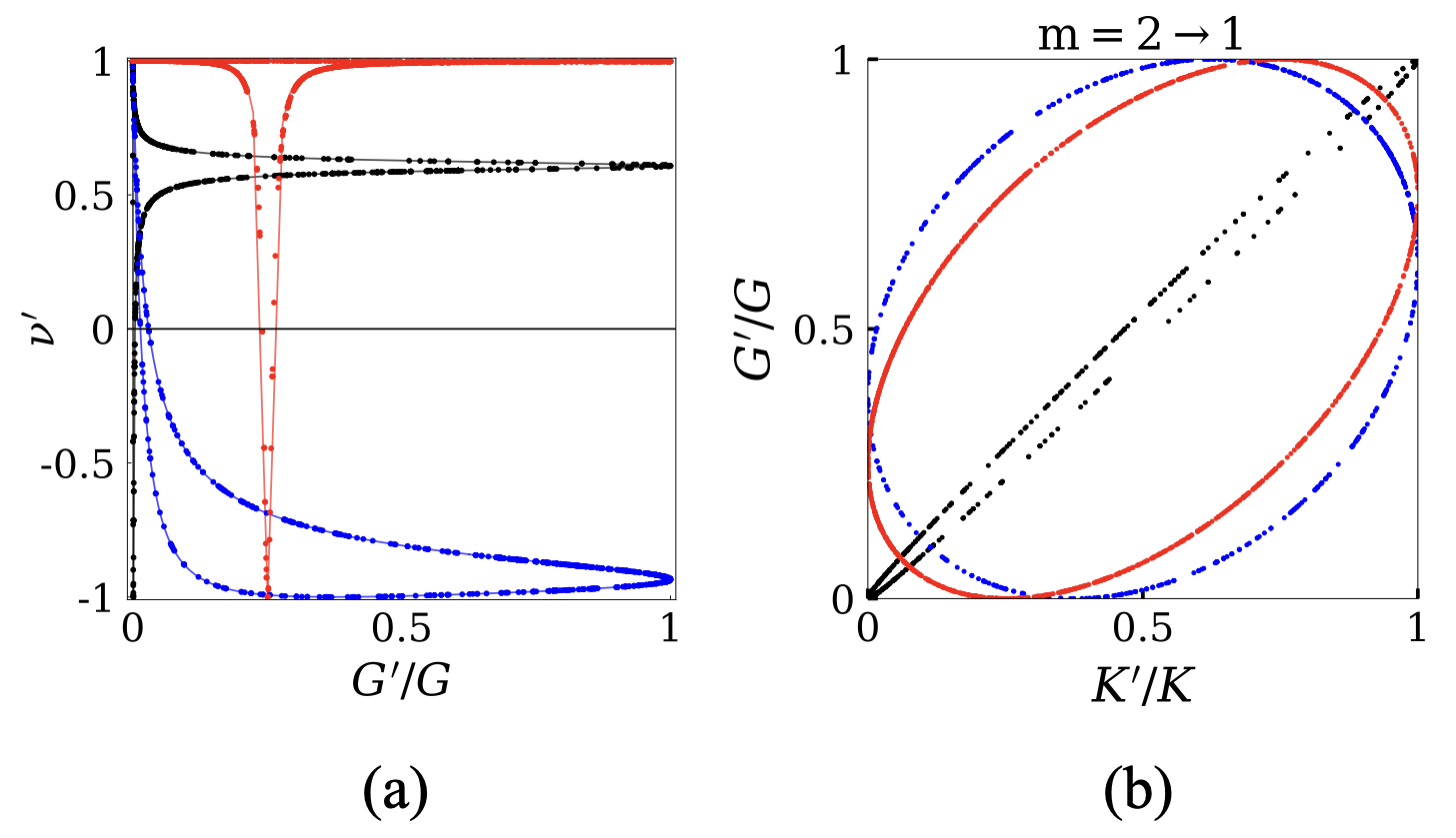}
\caption{(a) Poisson's ratio, $\nu'$, of three networks each with $m=1$ SSS and various values of $\eta$:
(black) a spring network generated by randomly removing stressed bonds from a Delaunay triangulation while imposing Hilbert's stability condition on each node with $\eta \approx 0.001$, $G/K \approx 0.244$, and $\nu \approx 0.610$ in the parent network;
(blue) an auxetic network with $\eta \approx 0.374$, and $G/K \approx 18.50$ and $\nu \approx -0.90$;
(red) a jammed network with $\eta \approx 0.251$, $G/K \approx 0.0006$, and $\nu \approx 0.998$.
The solid lines are drawn using Eq.~\ref{eq:ellipse} and Eq.~\ref{eq:PoissonsRatio} and data points are obtained by removing bonds and measuring the elastic moduli directly from the networks. In all three networks, there are bonds that change the sign of the Poisson's ratio when removed. (b) The characteristic  ellipses of the three networks shown in (a). Each color matches the color of its corresponding curve in (a).
}
\label{fig:PoissonsRatio}
\end{figure}

In Fig.~\ref{fig:PoissonsRatio}, we show the Poisson's ratios of three types of networks that are brought from $m=2$ SSS to $m=1$ SSS by removing a single bond ($m=2 \rightarrow 1$). As can be seen from the figure, in all three types of networks, there are bonds that can flip the sign of the Poisson's ratio when removed.
This is particularly unusual in the case of jamming (red), since jammed packings of soft particles are known to attain a finite bulk modulus while the shear modulus goes to zero at the onset of jamming~\cite{ellenbroek2009non}. 

In $2D$, $\nu'$ to the first order of $G'/K'$ is given by $1 - 2G'/K'$, implying that in systems such as jammed packings where usually $G' \ll K'$, the system remains non-auxetic when a bond is removed. However, this behavior breaks down when $K'$ is sufficiently small (this can occur even if $G \ll K$ in the parent system). In this regime, 
the system can only be made auxetic ($\nu' < 0$) when $K' \approx 0$ or equivalently $G'/G \approx \eta$ (from Eq.~\ref{eq:ellipse}), otherwise the Poisson’s ratio remains close to $1$. Fig.~\ref{fig:PoissonsRatio} shows an example of this regime (red curve) where the Poisson's ratio only deviates from values close to $1$ when $G'/G \approx \eta$. 
The small discrepancy between the data points and the theoretical solid red curve is due to the propagation of numerical errors.

In auxetic parent networks with $G > K$, the Poisson's ratio of the daughter networks $\nu'$ will typically remain negative. 
However, counter-intuitively, we observe that the further removal of certain bonds can indeed turn such an auxetic parent network into a non-auxetic daughter network with $G' < K'$ (see the blue curve in Fig.~\ref{fig:PoissonsRatio}).
We demonstrate that in elastic systems such as spring networks and jammed packings of soft harmonic spheres with only one state of self-stress, the elasticity of the system can be fully described by a single modulus. This is due to the functional dependence that emerges between the two independent moduli that exist at a higher distance from the onset of rigidity, when the number of states of self-stress is larger than one. 
In going from two states to one state of self-stress, a class of $m=2\to 1$ systems exist where removing a bond leads to strong changes in only one of the elastic moduli, almost decoupling the two. This is not usually possible in materials where all moduli are a consequence of the same micro-structure, and it is usually not possible with such minimal changes. On the other hand, there is a class of $m=2\to 1$ systems where removing a bond typically leads to a parallel weakening of all moduli (e.g. up to an order of magnitude reduction), but without making the system lose its rigidity. For any value of bulk or shear modulus in both types of systems, there are two possible values for the other modulus and we show that in moderately large systems, one can choose the resulting bulk or shear modulus, or equivalently a wide range of Poisson's ratios by removing a bond. This makes critically elastic systems a vastly tunable material in the case where single bond removal mechanisms can be implemented in an experimental system~\cite{rocks2017designing,reid2018auxetic}.
The results reported here have been tested on jammed networks and networks that are produced by selective pruning of bonds from a Delaunay triangulation, including networks that have a larger bulk modulus compared to shear modulus, and auxetic networks that have a significantly larger shear modulus than bulk modulus at $m=2$ SSS. One natural generalization would thus be to include randomly diluted networks where both moduli are typically of the same order and infinitesimal near the rigidity transition point~\cite{ellenbroek2015rigidity}.

We are grateful to Eric Corwin and Sascha Hilgenfeldt for inspiring conversations and for their substantial comments on the manuscript. This work has been supported by the Simons Foundation under grant No. 348126 to Sidney Nagel (VFH) and by National Science Foundation under grant DMS 1564468 to Michael Thorpe (MS).

\bibliography{main}

\clearpage

\appendix
\section{Bulk and Shear Strains}\label{bulkshearstrain}
Linear elasticity uses infinitesimal strain, $\bm{\epsilon}$, represented by a $d\times d$ deformation gradient matrix $\bm{\delta} = \mathbb{1} + \bm{\epsilon}$, where $\mathbb{1}$ is the identity matrix and $ |\bm{\epsilon}| \ll 1$. Applying $\bm{\delta}$ to the positions of the nodes in an elastic network ($\bm{r}'= \bm{\delta} \bm{r}$) results in an affine change in the length of bond $l$, which to linear order is given by~\cite{lubensky2015phonons}:
\begin{equation}\label{eq:DeltaStrain}
  \Delta_{l}
 = \ell_l \mathbf{n}_{l}^{T} \bm{\epsilon} \mathbf{n}_{l},
\end{equation}
where $\ell_l$ is the length of the bond before deformation is applied and $\mathbf{n}_{l}$ is a unit vector along the bond.

We focus on two types of deformation strains, namely compressional strain and shear strain, as described in the main text. In $2D$, these are defined as:
\begin{equation}
  (\bm{\epsilon}_{+})_{IJ} = \epsilon \delta_{IJ}, \quad (\bm{\epsilon}_{-})_{IJ} = \epsilon (-1)^{I+1} \delta_{IJ}.
\end{equation}
where $\delta_{IJ}$ is the Kronecker delta function, and $\bm{\epsilon}_{+}$ and $\bm{\epsilon}_{-}$ correspond to deformations probing the bulk and shear moduli, respectively. Indices $I$ and $J$ are $\in \{1,2\}$.
By plugging these strains into Eq.~\ref{eq:DeltaStrain}, we first obtain the bulk deformation vector given by:
\begin{equation}\label{}
  \Delta_{+,l}
  = \ell_l \mathbf{n}_{l}^{T} \bm{\epsilon}_{+} \mathbf{n}_{l}
  = \epsilon \ell_l \mathbf{n}_{l}^{T} \mathbf{n}_{l}
  = \epsilon \ell_l,
\end{equation}
for any bond $l$ (since $\mathbf{n}_{l}$ is a unit vector, $\mathbf{n}_{l}^{T} \mathbf{n}_{l} = 1$). For the shear response, deformations are applied in opposite directions and the change in length of bond $l$ is:
\begin{equation}\label{}
  \Delta_{-,l}
  = \ell_l \mathbf{n}_{l}^{T} \bm{\epsilon}_{-} \mathbf{n}_{l}
  = \epsilon \ell_l \sum_{i=1}^{d} (-1)^{i+1} \mathbf{n}_{l, i}^{2} = \epsilon \ell_l q_l.
\end{equation}
These two deformations can be written in vector form as $\bm{\Delta_{+}} = \epsilon \mathbf{L}$ and $ \bm{\Delta_{-}} = \epsilon \mathbf{Q} \mathbf{L} = \mathbf{Q} \bm{\Delta_{+}}$ where $\mathbf{Q}$ is a diagonal $N_c \times N_c$ matrix with diagonal elements $q_l$, and $\mathbf{L}$ is the vector of all bond lengths. Note that in calculating the bulk modulus, changes in the bond lengths are merely a function of the lengths before the bulk deformation is applied. In the case of shear modulus, however, the bond orientation plays a role through $q_l$. In $2D$, and for a bond that makes an angle $\theta_{l}$ with the $x-$axis, we have:
\begin{equation}
    q_l = \mathbf{n}_{l, 1}^{2} - \mathbf{n}_{l, 2}^{2} = \cos^{2}(\theta_{l}) - \sin^{2}(\theta_{l}) = \cos(2\theta_{l}).
\end{equation}

\section{The Ellipse Equation}\label{ellipse_eq_proof}
Suppose $(x,y)=(K'/K,G'/G)$ is a point on a curve that is defined by the parameter $\chi$:
\begin{align}
    \label{eq:param_appendix}
    x(\chi) &= \frac{(a_1+\chi a_2)^2}{(a_1^2+a_2^2)(1+\chi^2)}, \nonumber \\
    y(\chi) &= \frac{(b_1+\chi b_2)^2}{(b_1^2+b_2^2)(1+\chi^2)}.
\end{align}
If we define:
\begin{equation}
    \eta \equiv \frac{(a_1 b_2 -  a_2 b_1)^2}{(b_1^2+b_2^2)(a_1^2+a_2^2)},
\end{equation}
the following points will be on the curve:
\begin{align}\label{eq:ptson}
    \chi=-a_1/a_2 &:\quad (0,\eta) \nonumber \\
    \chi=-b_1/b_2&:\quad (\eta,0) \nonumber \\
    \chi=a_2/a_1&:\quad (1, 1-\eta) \nonumber \\
    \chi=b_2/b_1&:\quad (1-\eta, 1) \nonumber \\
    \chi=0 &:\quad \Big(\frac{a_1^2}{a_1^2+a_2^2}, \frac{b_1^2}{b_1^2+b_2^2}\Big).
\end{align}
Note that $x=K'/K$, $y=G'/G$, and $\eta$ are all bounded between $0$ and $1$. 
By plugging the above values into the most general form of a two dimensional conic section, we find the equation of an ellipse. 

Note that although $\eta$ assumes a specific value for a given network, these equations must be valid for any $\eta$. 
Since $(0,\eta)$ and $(\eta, 0)$ are both on the curve, $y=x$ is a symmetry line and parallel to one of the axes of the ellipse while the other axis is parallel to $y=-x$, since $(1-\eta, 1)$ and $(1, 1-\eta)$ are also on the curve. 
Therefore the ellipse is rotated at $\pm \frac{\pi}{4}$ with its center at $(1/2,1/2)$. The most general form of an ellipse with such characteristics is:
\begin{align}
    \left(\frac{x+y-1}{R_1}\right)^2 + \left(\frac{x-y}{R_2}\right)^2 = 2.
\end{align}
where $R_1$ and $R_2$ are the two axes of the ellipse.
By substituting any of the points in~\ref{eq:ptson}, we find:
\begin{equation}
    \eta^2 - \left( \frac{2 R_2^2}{R_1^2+R_2^2} \right) \eta + \left( \frac{R_2^2 (1-2R_1^2)}{R_1^2+R_2^2} \right) = 0
\end{equation}
which is a quadratic equation of $\eta$ with the following solutions:
\begin{equation}
\eta = \frac{R_2^2}{R_1^2+R_2^2} \pm \frac{R_1 R_2}{(R_1^2+R_2^2)} \sqrt{2R_2^2+2R_1^2-1}.
\end{equation}
However, $\eta$ has a single value for each network, hence this equation should have a double root and the discriminant must be zero ($2R_2^2+2R_1^2-1 = 0$). This means,
\begin{equation}
    \eta = \frac{R_2^2}{R_1^2+R_2^2} = 2R_2^2 = 1-2R_1^2,
\end{equation}
or:
\begin{align}
    R_1^2 &= \frac{1-\eta}{2}, \nonumber \\
    R_2^2 &= \frac{\eta}{2}.
\end{align}
Therefore points $(x,y)$ form the following ellipse:
\begin{equation}
    \left(\frac{x+y-1}{\sqrt{1-\eta}}\right)^2 + \left(\frac{x-y}{\sqrt{\eta}}\right)^2 = 1
\end{equation}
where for $ 0  <\eta < 1/2$, $R_1>R_2$ and for $1/2 <\eta < 1$, $R_1<R_2$. If $\eta=1/2$, $R_1 = R_2$ and the ellipse is a circle. For $\eta=0$ or $\eta=1$, the ellipse will reduce to a line.

Finally, the eccentricity of the ellipse for $\eta \geq 1/2$ is:
\begin{equation}
e=\sqrt{1 - \frac{R_1^2}{R_2^2}} = 
   \sqrt{\frac{2\eta - 1}{\eta}} = \sqrt{2 - \frac{1}{\eta}}
\end{equation}
and for $\eta < 1/2$ is:
\begin{equation}
e=\sqrt{1 - \frac{R_2^2}{R_1^2}} = 
   \sqrt{\frac{1 - 2\eta}{1-\eta}}.
\end{equation}

\section{Derivation of Arcsine Distribution of $\eta$}\label{Arcsine}

In this section, we analytically derive the distribution of $\eta = \sin^2\theta$, under the assumption that $\theta$ is uniformly distributed over $(0, \pi)$. This supports the empirical observation in Fig.~4 and justifies the U-shaped form of the $\eta$ distribution under spatial perturbations.

Let $\bm{\Delta}_{+} = a_1 \mathbf{s}_1 + a_2 \mathbf{s}_2$ and $\bm{\Delta}_{-} = b_1 \mathbf{s}_1 + b_2 \mathbf{s}_2$ be two vectors in a self-stress plane. We define the angle $\theta$ as the angle between $\bm{\Delta}_{+}$ and $\bm{\Delta}_{-}$. According to Eq.~8, $\eta$ can be represented as:

\[
\eta = \frac{\left\| \bm{\Delta}_{+} \times \bm{\Delta}_{-} \right\|^2}{\left\| \bm{\Delta}_{+} \right\|^2 \left\| \bm{\Delta}_{-} \right\|^2} = \sin^2 \theta
\]

Now assume that $\theta$ is a continuous random variable uniformly distributed over the interval $[0, \pi]$:
\[
\theta \sim \text{Uniform}(0, \pi), \quad \text{with PDF} \quad f_\theta(\theta) = \frac{1}{\pi}, \quad \text{for } 0 \leq \theta \leq \pi.
\]

Our goal is to derive the probability density function \( f_\eta(\eta) \) for the transformed variable \( \eta = \sin^2\theta \), where \( \eta \in (0,1) \).

The function \( \eta = \sin^2\theta \) is not one-to-one on the interval \( [0, \pi] \), but it is symmetric about \( \theta = \frac{\pi}{2} \). For each \( \eta \in (0,1) \), there are exactly two corresponding values of \( \theta \) in the domain:
\[
\theta_1 = \arcsin(\sqrt{\eta}), \quad \theta_2 = \pi - \arcsin(\sqrt{\eta}).
\]

To apply the change-of-variables technique, we compute the derivative of the inverse transformation:
\[
\frac{d\theta}{d\eta} = \frac{d}{d\eta} \arcsin(\sqrt{\eta}) = \frac{1}{2\sqrt{\eta(1 - \eta)}}.
\]

Using the standard formula for change of variables in probability, accounting for multiple branches:
\[
f_\eta(\eta) = \sum_{i=1}^{2} f_\theta(\theta_i) \left| \frac{d\theta_i}{d\eta} \right|.
\]

Since both $\theta_1$ and $\theta_2$ lie within the support of $\theta$, and the PDF of $\theta$ is constant:
\[
f_\eta(\eta) = \frac{1}{\pi} \cdot \left( \frac{1}{2\sqrt{\eta(1 - \eta)}} + \frac{1}{2\sqrt{\eta(1 - \eta)}} \right) = \frac{1}{\pi \sqrt{\eta(1 - \eta)}}.
\]

Therefore, the probability density function of the transformed variable \( \eta = \sin^2\theta \), when \( \theta \sim \text{Uniform}(0, \pi) \), is:
\[
f_\eta(\eta) = \frac{1}{\pi \sqrt{\eta(1 - \eta)}}, \quad \text{for } \eta \in (0,1),
\]
which is known as the \emph{arcsine distribution}.

\end{document}